\def\year{2022}\relax
\documentclass[letterpaper]{article} 
\usepackage{aaai22}  
\usepackage{times}  
\usepackage{helvet}  
\usepackage{courier}  
\usepackage[hyphens]{url}  
\usepackage{graphicx} 
\usepackage{natbib}  
\usepackage{caption} 
\usepackage{todonotes}
\DeclareCaptionStyle{ruled}{labelfont=normalfont,labelsep=colon,strut=off} 
\usepackage{algorithm}
\usepackage{algorithmic}
\usepackage{amsmath}
\usepackage{booktabs}
\usepackage{multirow}
\usepackage{makecell}
\usepackage{subcaption}
\usepackage{newfloat}
\usepackage{listings}
\floatstyle{ruled}
\newfloat{listing}{tb}{lst}{}
\floatname{listing}{Listing}
\title{Moral Values Underpinning COVID-19 Online Communication Patterns}

\author {
Julie Jiang,
Luca Luceri, 
Emilio Ferrara 
}
\affiliations {
Information Sciences Institute, University of Southern California, Marine del Rey, CA, USA \\
juliej@isi.edu, lluceri@isi.edu, ferrarae@isi.edu
}

\begin{document}

\maketitle

\begin{abstract}
The COVID-19 pandemic has triggered profound societal changes, extending beyond its health impacts to the moralization of behaviors. Leveraging insights from moral psychology, this study delves into the moral fabric shaping online discussions surrounding COVID-19 over a span of nearly two years. Our investigation identifies four distinct user groups characterized by differences in morality, political ideology, and communication styles. We underscore the intricate relationship between moral differences and political ideologies, revealing a nuanced picture where moral orientations do not rigidly separate users politically. Furthermore, we uncover patterns of moral homophily within the social network, highlighting the existence of one potential moral echo chamber. Analyzing the moral themes embedded in messages, we observe that messages featuring moral foundations not typically favored by their authors, as well as those incorporating multiple moral foundations, resonate more effectively with out-group members. This research contributes valuable insights into the complex interplay between moral foundations, communication dynamics, and network structures on Twitter.

\end{abstract}
\section{Introduction}

The seismic upheaval brought about by the COVID-19 pandemic transcended more than just illness and death. As almost unanimously the single biggest event of the 21st century yet, it catalyzed a series of important changes disrupting the lives of nearly everyone across the world. 
Actions that once constituted ordinary conduct, such as breathing freely without face coverings, became morally laden with accusations of selfishness and harm \cite{latimes2020}. Work swiftly transitioned from in-person to remote, changing normative expectations of work \cite{pew2022normal}. Research consistently underscores the intertwining of moral judgments with decisions pertaining to health-related behaviors, such as the intricate linkages between morality and vaccine hesitancy \cite{schmidtke2022evaluating,reimer2022moral}.

In this paper, we use moral psychology to explain what shapes online conversations surrounding COVID-19. The Moral Foundation Theory (MFT) was created to explain the origins and elements of human moral reasoning. It consists of five main foundations: \textit{care/harm}, \textit{fairness/cheating}, \textit{loyalty/betrayal}, \textit{authority/subversion}, and \textit{purity/degradation} \cite{haidt2004intuitive,graham2013moral}. These foundations serve as the compass guiding our moral compass, shaping how we perceive and engage with the world around us. While individuals may embody varying degrees of these foundations, the misalignment of moral orientations between individuals often leads to disagreement, discord, and moral conflicts \cite{haidt2012righteous}. More recently, computational social scientists have also uncovered patterns of moral foundation homophily in social networks. For instance, \citet{dehghani2016purity} highlighted the predilection for social networks to exhibit \textit{purity} homophily—a preference for network connections who share similar purity moral values.

The burgeoning research on moral psychology has led to the discovery of its link with political ideology \cite{haidt2007morality,graham2009liberals,graham2012moral,haidt2012righteous,feinberg2019moral,hatemi2019ideology}. Liberals tend to base their morals on the two \textit{individualizing} foundations--\textit{care} and \textit{fairness}--while conservatives tend to base their morals on all five original foundations, including the \textit{binding} foundations of \textit{loyalty}, \textit{authority}, and \textit{purity}. As a result, the political divide that has long persisted in American politics can be viewed as a clash between fundamentally different sets of moralities \cite{haidt2012righteous,feinberg2019moral}.

Building upon these premises and drawing from the moralization of COVID-19, this study delves into the landscape of discussions about the pandemic on Twitter. We navigate questions poised at the intersection of morality, communication patterns, and network dynamics, seeking to unravel the moral fabric underpinning discourse amidst a global crisis. Specifically, we pose the following research questions:
\begin{itemize}
\item \textbf{RQ1}: What distinct user groups emerge in Twitter discussions regarding COVID-19 concerning moral frameworks and network interactions?
\item \textbf{RQ2}: Do communication patterns among these user-defined moral groups reflect tendencies of moral homophily within the network?
\item \textbf{RQ3}: Which moral themes within messages exhibit a more effective resonance within in-group and out-group interactions?
\end{itemize}

Through this exploration, we aim to uncover the intricate interplay between moral foundations, communication dynamics, and network structures within the context of COVID-19 discourse on social media. Our summary of contributions is as follows. First, we discover four main groups of users with vastly different moral priorities and political partisanship. We paint a nuanced picture of the relationship between morality and political ideology, demonstrating that moral orientations do not rigidly separate users across the political spectrum. We also find that most user groups exhibit group-based homophily--the tendency to communicate with in-group members. One group of users (group IV) who are primarily right-leaning users with \textit{fairness} and \textit{authority} moral foundations also exhibit tendencies to only communicate with in-group members, a trend that may suggest moral echo chambers. Finally, we find that messages with moral foundations that are not typically favored by their authors and messages with moral pluralism tend to resonate better with out-group members. We conclude with insights into user group dynamics and communication patterns, emphasizing the importance of moral diversity for fostering effective discourse across diverse perspectives.

\section{Background and Related Work}

\subsection{The Moral Foundation Theory}
The Moral Foundation Theory was proposed to explain variations in human moral reasoning. The original MFT consists of the following five main foundations \cite{haidt2004intuitive,graham2013moral}:

\begin{enumerate}
    \item \textit{Care/harm}: This foundation relates to our ability to feel empathy and compassion for others and our willingness to alleviate their suffering. It emphasizes the importance of caring for and protecting others, especially those who are vulnerable.
    \item \textit{Fairness/cheating}: This foundation emphasizes the importance of justice, equality, and fairness in our interactions with others. It focuses on the belief that everyone should be treated fairly and equally.
    \item \textit{Authority/subversion}: This foundation relates to our respect for authority, hierarchy, and tradition. It emphasizes the importance of following rules and obeying authority figures.
    \item \textit{Loyalty/betrayal}: This foundation underlies the sense of belonging to a group and the importance of showing loyalty and allegiance to that group. It emphasizes the importance of being patriotic and self-sacrificing for the betterment of the group.

    \item \textit{Purity/degradation}: Also known as \textit{sanctity/degradation},\footnote{We consider ``sanctity'' and ``purity'' to be interchangeable terms when drawing from literature, but use ``purity'' throughout this paper to maintain consistency.} this foundation relates to our sense of purity and cleanliness and the importance of self-control to avoid impure or degrading actions. It emphasizes the importance of protecting sanctity and avoiding contamination. 
\end{enumerate}


There is also a sixth foundation, \textit{liberty/oppression}, proposed in \citet{haidt2012righteous}. This foundation relates to our belief in individual freedom and autonomy, as well as our opposition to oppression and coercion. However, since this foundation was often omitted in related research in MFT detection from texts \cite{guo2023data,johnson2018classification,rojecki2021moral}, we also do not consider this foundation in this work.

\subsection{Morality and Politics}

There is substantial research on how morality binds and divides users by political orientation \cite{haidt2007morality,graham2009liberals}. Users with liberal ideology typically reflect higher individualizing moral foundations of \textit{care} and \textit{fairness}, foundations that support the rights and welfare of individuals, whereas conservative ideology typically endorses all five moral foundations equally. It follows that, in comparison with the liberals, the conservatives have comparatively binding moral foundations of \textit{authority}, \textit{loyalty}, and \textit{purity}, also known as the three binding moral foundations that promote the rights and welfare of the group and the institution. \citet{koleva2012tracing} further exemplified this by examining 20 politically salient issues such as abortion and immigration. They found that the five moral foundations, in particular \textit{purity}, are better predictors of issue-specific opinion above ideology and demographic features. This difference in moral attitude has been attributed to growing political polarization, where users on both ends of the political spectrum are unable to resonate with each other due to moral incongruence \cite{haidt2007morality} and overexaggerate their differences \cite{graham2012moral}. Techniques of moral-based reframing has thus been proposed as a way to bridge the political divide \cite{feinberg2019moral}.

\subsection{Moral Homophily}
Another research direction is on how moral values explain the connection and formation of communities. Computational social scientists have long recognized the powers of social network homophily, that we tend to be drawn to people we are similar to \cite{mcpherson2001birds,kossinets2009origins}. As such, it stands to reason that moral foundations, just like age, religion, and education, may also play a role in shaping our social network attachments. In a large-scale analysis of social network data on the US government shutdown, \citet{dehghani2016purity} found that the \textit{purity} foundation, but not other moral foundations, predicts network ties. In another study comparing multilingual tweets explicitly mentioning morality in English and Japanese, \citet{singh2021morality} found that the \textit{care}, \textit{authority} and \textit{purity} foundations are homophilous in English, while the \textit{loyalty}, \textit{authority} and \textit{purity} foundations are homophilous in Japanese. We theorize that for our topic of COVID-19, some moral foundations may exhibit network homophily, potentially informing our moral-based understanding of online conversations.

\subsection{Morality and COVID-19}


Recent research has shown that COVID-19 brought about many divisive behaviors and controversial opinions that may be rooted in moral differences. Those with a higher \textit{care} disposition are found to be more likely to follow health recommendations \cite{diaz2022reactance}. A similar study by \citet{chan2021moral} also found that higher \textit{care} and \textit{fairness} predicts compliance with the health strategies of staying at home, wearing masks, and social distancing, while \textit{purity} predicts non-compliance with wearing masks and social distancing. In the face of disease threats, \citet{ekici2021deciding} found that \textit{fairness}, \textit{care} and \textit{purity} were the most important moral foundations that predicted people's acceptability of moral transgressions. In a study of tweets on the COVID-19 mask mandate, \citet{mejova2023authority} found that \textit{authority} and \textit{purity} were associated with anti-masking sentiment while \textit{fairness} and \textit{loyalty} were associated with pro-masking sentiment. There is also a deemphasis on the \textit{care} foundation following the mask mandate.


By 2021, arguments loaded with moral judgments both for and against the COVID-19 vaccination take center stage. In a study based in Great Britain, \citet{schmidtke2022evaluating} found that vaccine hesitancy is associated with higher moral needs of \textit{authority},  \textit{liberty}, and \textit{purity} and less need of \textit{care}. Moral foundations were also shown to be good predictors of country-level vaccination rates in the US: \textit{purity} predicted lower vaccination rates, whereas \textit{fairness} and \textit{loyalty} predicted higher vaccination rates \cite{reimer2022moral}. Analyzing the sentiment in tweets related to vaccination, \citet{pachec2022holistic} found that pro-vaccination tweets carried more \textit{care} morals, while anti-vaccination tweets carried more \textit{liberty} morals. In another social media platform, \citet{beiro2023moral} found that Facebook posts on vaccination surrounds moralities of \textit{liberty}, \textit{care}, and \textit{authority}, with pro-vaccination users identifying more with \textit{authority} and anti-vaccination users identifying more with \textit{liberty}. The debate on vaccination is also linked to partisanship. Liberals discuss COVID vaccination on Twitter with more emphasis on the moral virtues of \textit{care}, \textit{fairness}, \textit{liberty}, and \textit{authority}, whereas conservatives leaned into the vices of \textit{oppression} and \textit{harm} \cite{borghouts2023understanding}.


Besides being moralized, COVID-19 was also politicized, which may have led to the formation of online political echo chambers \cite{jiang2020political,jiang2021social}. Echo chambers are harmful to online ecosystems as communication is politically segregated, contributing to radicalism and extremism \cite{o2015echo}. Investigating both the political and moral nature of COVID-19, \citet{bruchmann2022moral} reveals that binding and individualizing moral foundations can explain partisan differences in attitudes: conservatives see vaccine hesitancy as more permissible and self-report fewer prevention behaviors. In another analysis of COVID-19 tweets, \citet{rao2023pandemic} found that conservatives, compared to liberals, employed more moral vices than virtues in their languages. Several works also tested the promising direction of moral reframing. \citet{luttrell2023advocating} studied using ideology-matched arguments for wearing masks, which they found to work well with liberals but not conservatives. \citet{kaplan2023moral} also conducted a moral reframing study on masks targeted at conservatives and showed success in messages advocating the use of masks with the \textit{loyalty} moral foundation.

\section{Data}

\begin{figure}
    \centering
    \includegraphics[width=\linewidth]{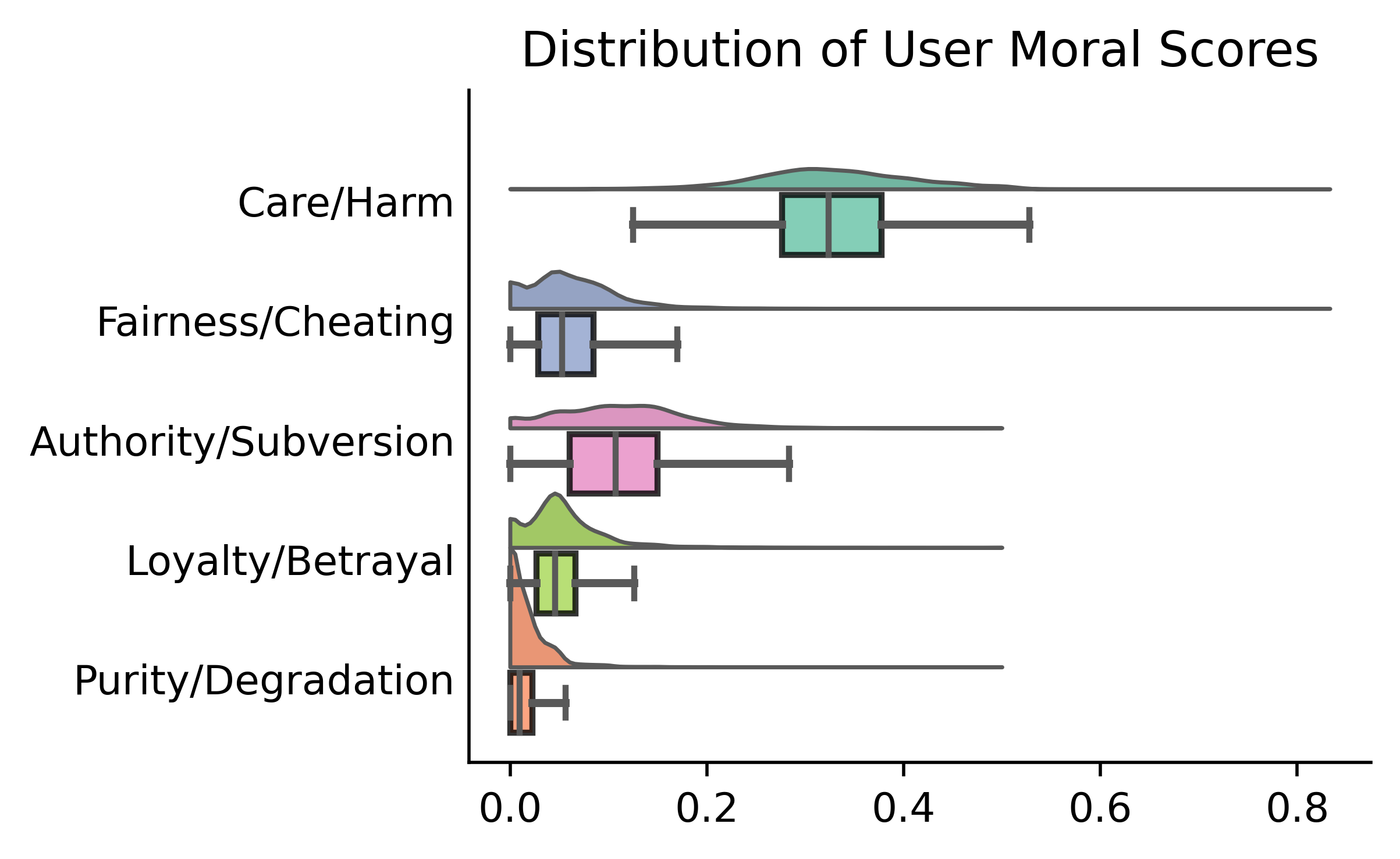}
    \caption{Distribution of raw user moral scores. }
    \label{fig:user_mf_dist}
\end{figure}


For this research, we use a large dataset of real-time COVID-19 tweets collected by \citet{chen2020tracking}, retrieved based on a set of curated keywords related to COVID-19. We use a subset of the data spanning the beginning of February 2020 to the end of October 2021 for 21 full months. We then take a longitudinal set of active users who tweeted at least 10 tweets containing moral values in any given month, which means they each contain at least 10 data points of moral foundation values. We describe how we detect moral values in the Methods section below. This procedure resulted in a large dataset of 2 million users and 253 million tweets. To analyze communication patterns, we also create retweet and mention networks. Using users as nodes, we build the retweet network by connecting users who retweet one another. Similarly, we build the mention network by connecting users who do not retweet but rather quote (retweet with additional comment) or mention (`@’) another user. We draw the distinction between the retweet and mention network as there may be different underlying motivations for each action. Retweeting usually implies endorsement \cite{boyd2010tweet,metaxas2015retweets} while mentioning could be used to endorse or criticize \cite{hemsley2018tweeting}. When identifying retweets or mentions between two target users, we use all available tweets, including those that do not have any identifiable moral foundations. The edges are weighted by the number of times one user retweets or mentions another. To allow for efficient computation, we create separate networks for each month of our dataset and aggregate the results if necessary. On average, we have 90,000 nodes, 1 million retweet edges, and 1.3 mention edges in our monthly networks.


\begin{table*}[]
    \centering
    \begin{tabular}{ll}
    \toprule
    \textbf{Tweet} & \textbf{Morals}\\
    \midrule
    \textit{The vaccine won't harm you if you are in one of the approved groups, but it will protect you.} & Care, Harm\\
    \midrule 
    \textit{New Zealand just announced it will provide the new Covid vaccine to any New Zealander who wants} &  \multirow{2}{*}{Fairness, Authority}\\
    \textit{it--free of charge. They're also making the vaccine available to all their Pacific island neighbors.}\\
    \midrule 
    \textit{It's up to us to slow the spread, save lives, \& keep businesses open. We have to work together.} & Care, Loyalty\\
    \midrule 
    \textit{I refuse to take this vaccination! It goes against my religious beliefs!} & Purity \\%
    \midrule 
    \textit{Revolting. The fact that XXX got the vaccine before healthcare workers and first responders.} & Degradation\\
     \bottomrule
    \end{tabular}
    \caption{Hypothetical tweets, adapted from real ones, that contain detected moral foundations values. Some tweets only contain the virtue or the vice of a foundation, and some tweets contain both.}
    \label{tab:eg_tweets}
\end{table*}

\section{Method}
\subsection{Detecting the Morality of Tweets and Users}
The moral labels of the tweets in this dataset are detected using the MFT detector developed recently by \citet{guo2023data}. This model adopts a data fusion technique to account for the fundamental shifts in morality based on the topics of the dataset. It is based on a pre-trained BERT model \cite{bert} that was fine-tuned on three different Twitter datasets annotated for morality, including one that was specifically on COVID \cite{rojecki2021moral}. This MFT detector predicts the presence of the 10 moral values---each foundation contains two opposite moral values for the virtue and the vice---in a multi-label manner for every tweet, with 1 indicating the presence of a moral value and 0 indicating the absence of it. We present some example tweets with moral foundation labels in Table \ref{tab:eg_tweets}. We then aggregate the virtues and vices of each foundation into one label. If a tweet has a score of 1 for the \textit{care} dimension but a score of 0 for the \textit{harm} dimension, then it has a combined score of 0.5 (1 / 2) in the \textit{care/harm} foundation. We choose to do this because the morality detector does not detect attitude but rather explicitly expression in either the virtue (e.g., \textit{care}) or the vice (e.g., \textit{harm}), both of which reflect a moral disposition in that foundation. The vast majority (87\%) of tweets contain only a single moral foundation. 13\% of the tweets contain two moral foundations, of which the most popular combination is \textit{care} and \textit{authority}. Tweets with 3 or 4 moral foundations make up less than 1\% of the total, and no tweets have all 5 moral foundations.

We also compute morality scores at the user level. For each user, we calculate the average moral score of each foundation based on all the tweets by that user. These five scores represent the user's morality profile. The distribution of the aggregated user morality scores is shown in Figure \ref{fig:user_mf_dist}. The \textit{care} foundation is the most frequently utilized foundation, followed by \textit{authority}. \textit{Purity} is the least utilized foundation. To maintain cross-comparison of each morality, we standardize (z-scores) the scores in each foundation to have a mean score of 0 and a standard deviation of 1. 

\subsection{Detecting User Groups Using MFT and Twitter Activity} 
In pursuit of our research goal, the communication dynamics among users with various moralities, we set out to find salient groups of users in our dataset based on their moral foundations and their communication preferences. For this task, we leverage Social-LLM \cite{jiang2023socialllm}, a social network user detection method that combines language features and network features using theories of homophily. Social-LLM is an unsupervised user representation method that learns user embeddings such that two users who share a retweet and mention a connection have embeddings that are similar to each other, as measured by cosine similarity. On a sample of our original dataset, Social-LLM user representations were shown to work well in predicting individual user moral foundations based on user metadata features and language embeddings of their profile descriptions. Building on the success shown in \citet{jiang2023socialllm}, we learn a set of user representations using network features, profile descriptions, user metadata features, and moral foundation scores by re-training Social-LLM on their sampled dataset. We use the best hyperparameter configuration, learning 128-dimensional embeddings using directed retweet and mention edges and a base language model of SBERT-MPNet. Then, we apply the model to our dataset to obtain user embeddings for our 2 million users. 

The learned user embeddings contain important cues about users' moral foundation values and Twitter activities in the form of social network interactions, profile descriptions, and other user metadata features. Then, we employ the $k$-means clustering algorithm on the embeddings to uncover distinct user groups. After experimenting with cluster numbers ranging from 2 to 10, we select $k=4$ as the most appropriate number of clusters using the elbow method based on the inertia and the silhouette method. These four distinct user groups, encapsulating the distinct moral orientations and Twitter behaviors within our dataset, warrant further exploration and analysis.

\subsection{Detecting User Partisanship}
As morality is often linked to differences in political partisanship \cite{graham2009liberals,koleva2012tracing}, and because numerous studies on COVID have shown that user opinions are politically divided \cite{jiang2020political,rao2023pandemic}, we also want to capture users' political partisanship in this study. To this end, we utilize another Social-LLM model trained for user political leaning detection \cite{retweetbert,jiang2023socialllm}. Compared to other political leaning detection methods on Twitter, Social-LLM has the advantage of learning crucial cues from not only the textual features of the tweets but also social network interaction features. The latter is particularly preferable since Twitter users are often politically segregated, especially on the topic of COVID-19 \cite{jiang2020political,jiang2021social}. Social-LLM works by leveraging user profile description similarity and network homophily. Applying the Social-LLM model for political leaning detection on COVID-19 datasets, we obtain political-leaning labels for every user in our dataset. Similar to prior work \cite{jiang2021social}, we bin the scores into quintiles to adjust for the left bias. Users falling within the 0-20\% range are labeled as very left-leaning, those in the 20-40\% range as left-leaning, the middle 40-60\% as moderate, the 60-80\% as right-leaning, and those in the 80-100\% range as very right-leaning.


\begin{figure}
    \centering
    \includegraphics[width=\linewidth]{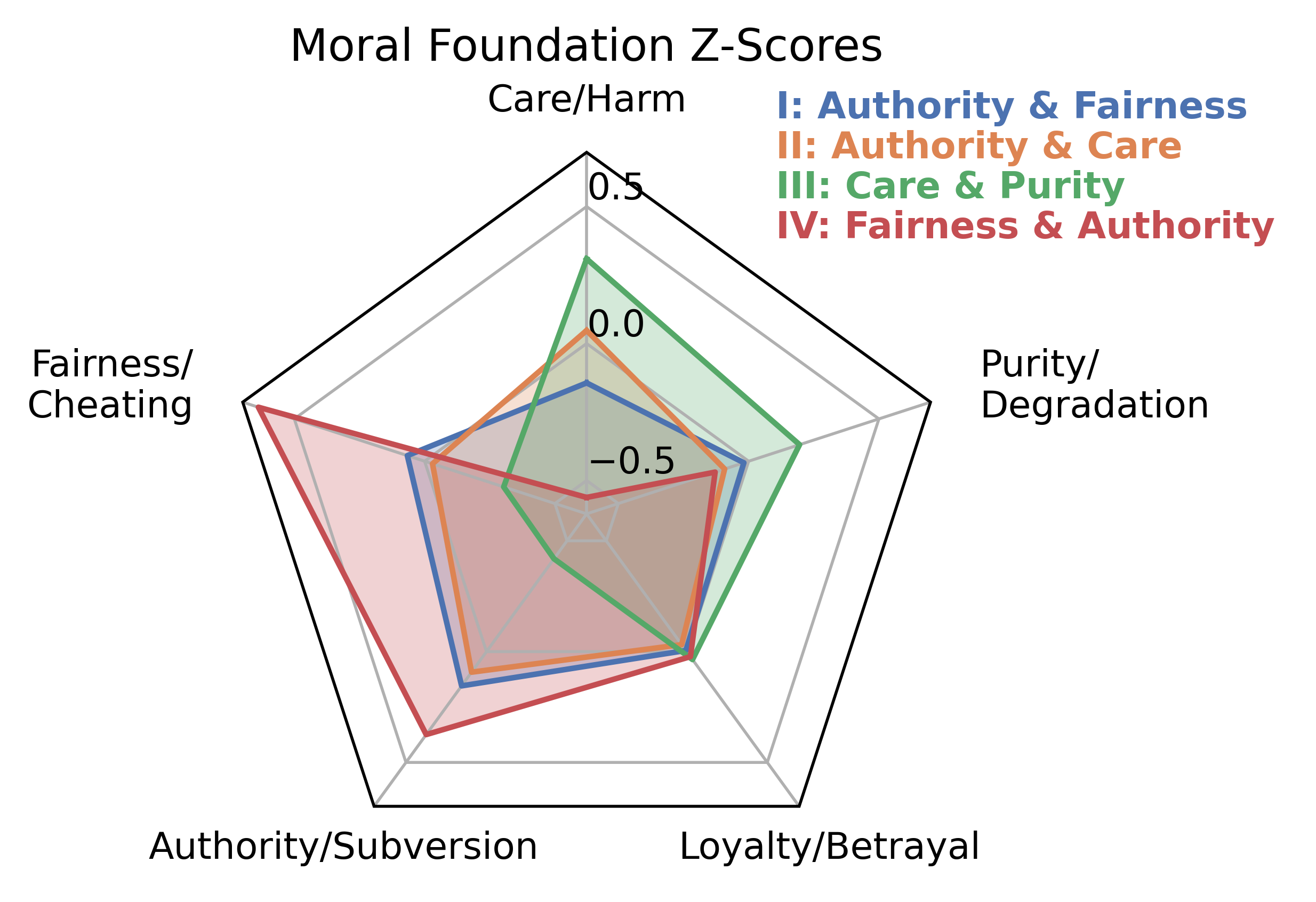}
    \caption{The average moral z-scores of each foundation for the four user groups.}
    \label{fig:user_clusters}
\end{figure}

\section{RQ1: User Groups by Morality}
\begin{figure}
    \centering
    \includegraphics[width=\linewidth]{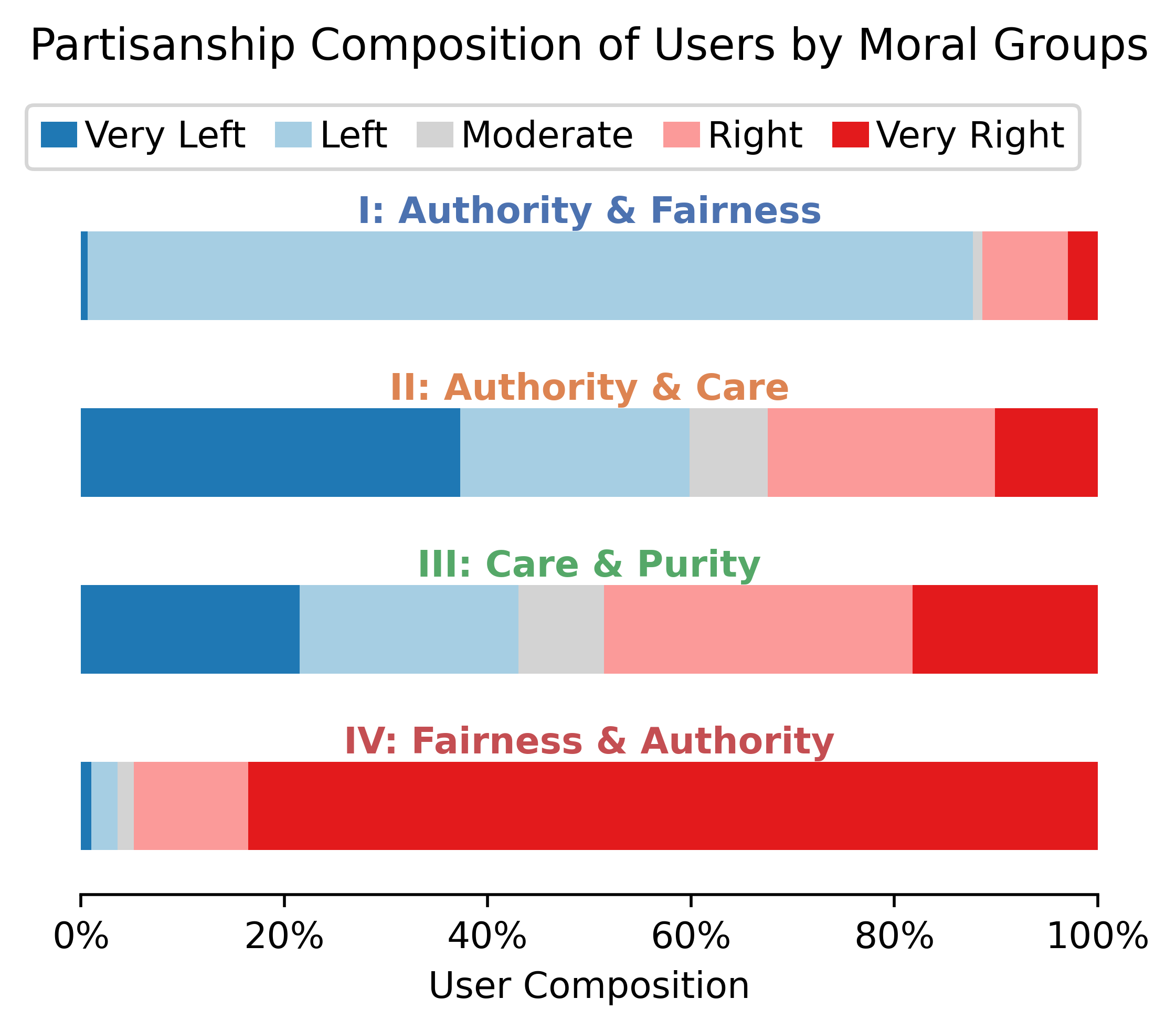}
    \caption{Partisanship breakdown of the four user groups. Blue bars represent left-leaning users and red bars represent right-leaning users}
    \label{fig:moral_poli_bar}
\end{figure}

\begin{figure*}
    \centering
    \includegraphics[width=\linewidth]{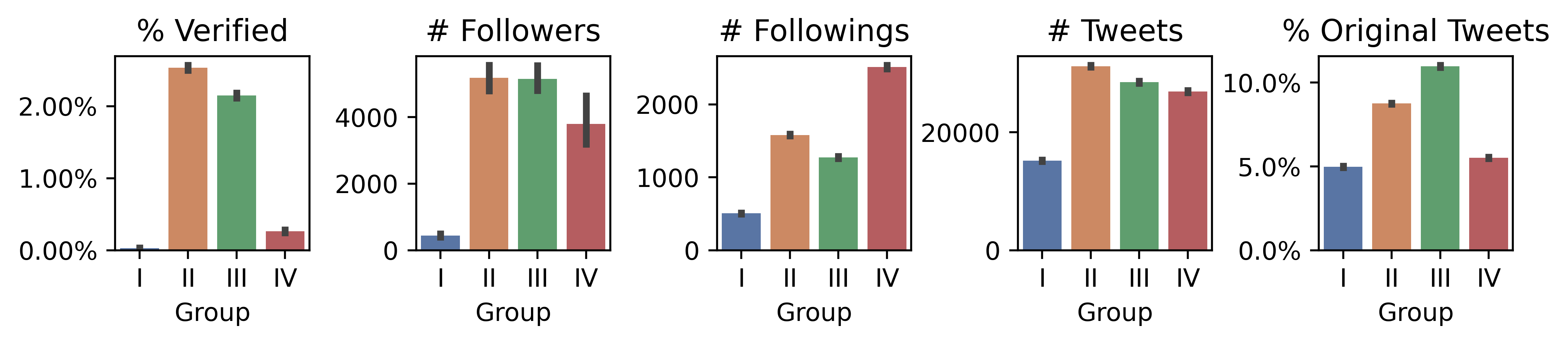}
    \caption{Distribution of the user metadata features for the four user groups.}
    \label{fig:user_meta}
\end{figure*}

We first answer the research question: what are the characteristics of the groups of users tweeting about COVID-19 based on their moral values, Twitter profiles, and network interactions? We show the average moral scores of each group in Figure \ref{fig:user_clusters}, a political partisanship breakdown in Figure \ref{fig:moral_poli_bar}, and some key user metadata statistics in Figure \ref{fig:user_meta}. Below, we discuss the characteristics of each group.

\paragraph{Group I: Authority \& Fairness.} This group contains 493,000 users, representing 23\% of all users in our dataset. These users reflect a relatively stronger morality in the \textit{authority} ($z=0.15$) and \textit{fairness} foundations ($z=0.07$). Its scores in the \textit{care} foundation are below average ($z=-0.14$). Considering user partisanship, we find that this group is predominantly occupied by left-leaning users: 87\% of its users have a left-leaning political score. However, it does not have many users that are extremely left-leaning.

We also note that the metadata features of group I users stand out in several important ways. They have the fewest number of verified users. On average, they have the least number of followers and followings, and they also posted substantially fewer posts. The proportion of original tweets of the tweets they did publish is also the lowest. In sum, Group I users appear to be the least active user group and most likely do not have as many influential or popular users as the other groups. As Twitter users reflect an overall left-bias \cite{jiang2021social}, we theorize that Group I users are the average Twitter users with low influence.

\paragraph{Group II: Authority \& Care.} With 800,000, or 37\% of users, this group is the largest user group in our dataset. Its users are characterized by higher moral scores in the \textit{authority} ($z=0.09$) and \textit{care} foundations $(z=0.05$). It scores comparatively lower in the \textit{purity} foundation ($z=-0.09$). In terms of partisanship, this group has a good balance of left- and right-leaning users, although it has a substantial amount of far-left users (38\%),

\paragraph{Group III: Care \& Purity.} Third group contains 603,000 (28\%) users and exhibits stronger than average morality in the \textit{care} ($z=0.30$) and \textit{purity} ($z=0.19$) foundations, along with a weaker than average morality in the \textit{authority} ($z=-0.42$) and \textit{fairness} ($z=-0.30$) foundations. It has a nearly perfect representation of users across all political leanings, comprising 48\% right-leaning (incl. far-right) and 44\% left-leaning (incl. far-left) users.

\paragraph{Group IV: Fairness \& Authority.} The fourth and final group is the smallest user group, made up of only 274,000 users, or 13\% of our use base. This group is characterized by stronger than average foundations of \textit{fairness} ($z=0.64$) and \textit{authority} ($z=0.37$), coupled with weaker than average foundations of \textit{care} ($z=-0.56$) and \textit{purity} ($z=-0.13$). This group is also made up of predominantly right-leaning users (95\%). Users who are far-right occupy 84\% of this group alone.

We note that the characteristics of these groups are discussed in relation to each other, not in absolute terms. That is, for example, Group I users display lower \textit{care} moral values than Group III users, but that does not mean Group I users don't utilize \textit{care} morality. Of the four user groups, we see that Groups III and IV users have very strong preferences in some moral foundations, whereas Group I and Group II have more modest preferences. Further, We note that Group III and Group IV users have almost polar opposite moral foundation inclinations. While Group III prefers \textit{care} and \textit{purity}, these are exactly the two moral foundations that are least utilized by Group IV, and vice versa for Group IV's preferred morality of \textit{fairness} and \textit{authority}. Additionally, though both Group I and Group IV prefer the foundations \textit{authority} and \textit{fairness}, they differ considerably in what foundations they don't prefer, the strength of their morality, and their political partisanship breakdown. Finally, while most of the five moral foundations are prominently favored or unfavored by at least one of the four user groups, the \textit{loyalty} foundation usage is used almost indistinguishable among the user groups.

\subsection{User Visualization} 
\begin{figure}
    \centering
    \includegraphics[width=\linewidth]{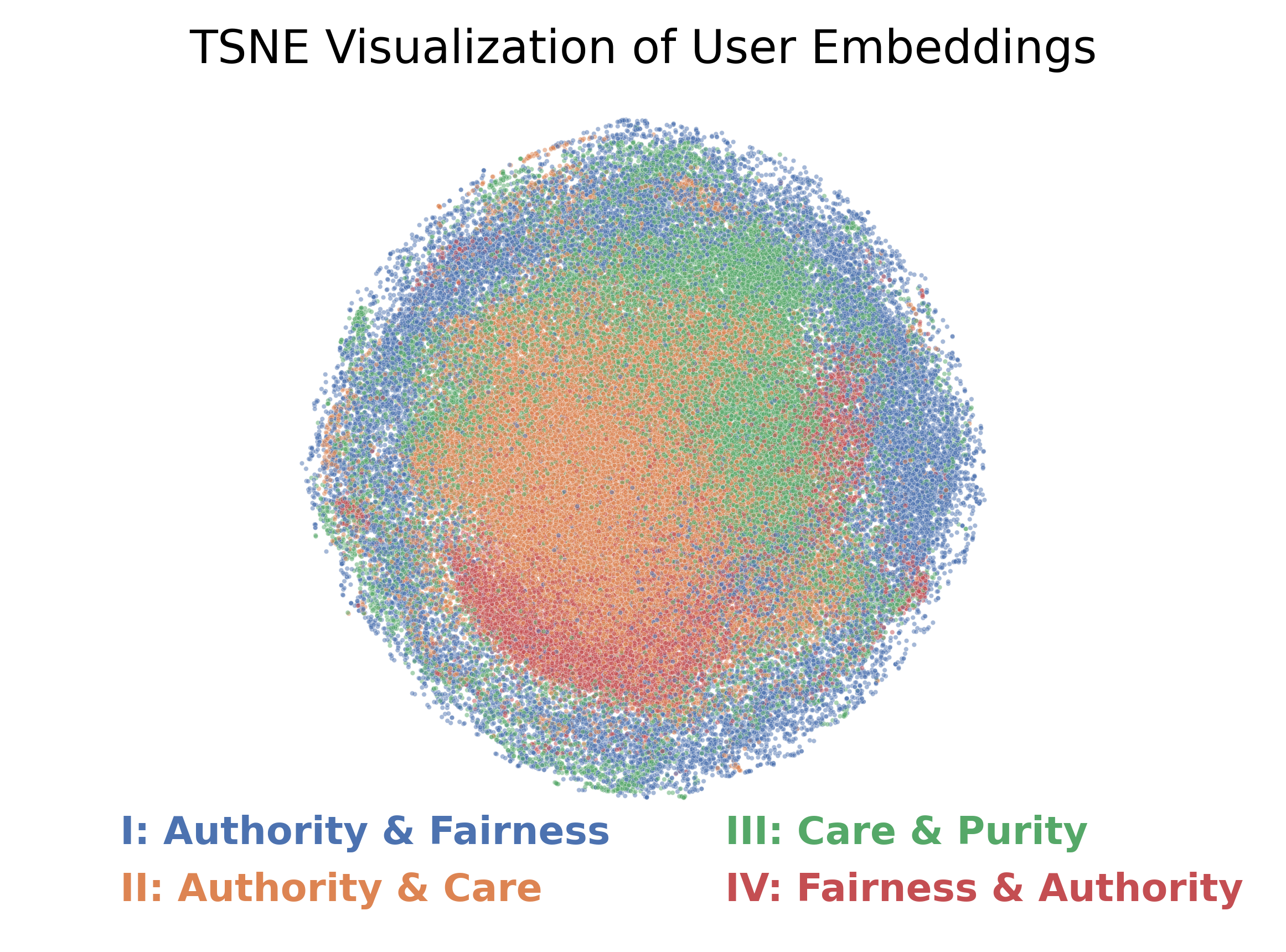}
    \caption{TSNE visualization of 100,000 sampled user embeddings of the four user groups.}
    \label{fig:tsne}
\end{figure}

In Figure \ref{fig:tsne}, we present a TSNE \cite{tsne} visualization of 100,000 sampled user embeddings in each group. TSNE is a popular dimension-reduction technique of high-dimensional data points to reveal structural proximity among users in each group. This plot shows a good separation of every user group, with user groups II, III, and IV forming visible clusters. This may indicate that these groups form homophilous communication bubbles. However, Group I users form a circular ring enclosing other user embeddings.  As we will see in the next section, this appears to be because users in this group do not preferentially communicate with in-group members but rather interact equally with all users.

\section{RQ2: Moral Homophily}
In this section, we continue our analysis by examining whether there is a communication homophily in terms of user moral typology. This can shed light on whether communication is barred among people who do not share similar moralities, which could lead to ineffective discussions and sowing dissent online.

\subsection{Homophily of Users}

The purpose of this section is to investigate moral homophily at the group level. However, as a preliminary analysis, we first see whether there is moral homophily in individual moral foundation values at the user level. That is, do users share similar moral foundation values as their network? To investigate this, we employ the network assortativity method \cite{assortativity}, which leverages the Pearson correlation coefficient to measure how similar two sets of nodes' attributes are, given that each pair of nodes is connected by an edge. Using the standardized moral scores as the users' node attributes and the retweet or mention interaction as edges, we compute the assortativity values of every moral foundation on a monthly basis. 

\begin{figure}
    \centering
    \includegraphics[width=\linewidth]{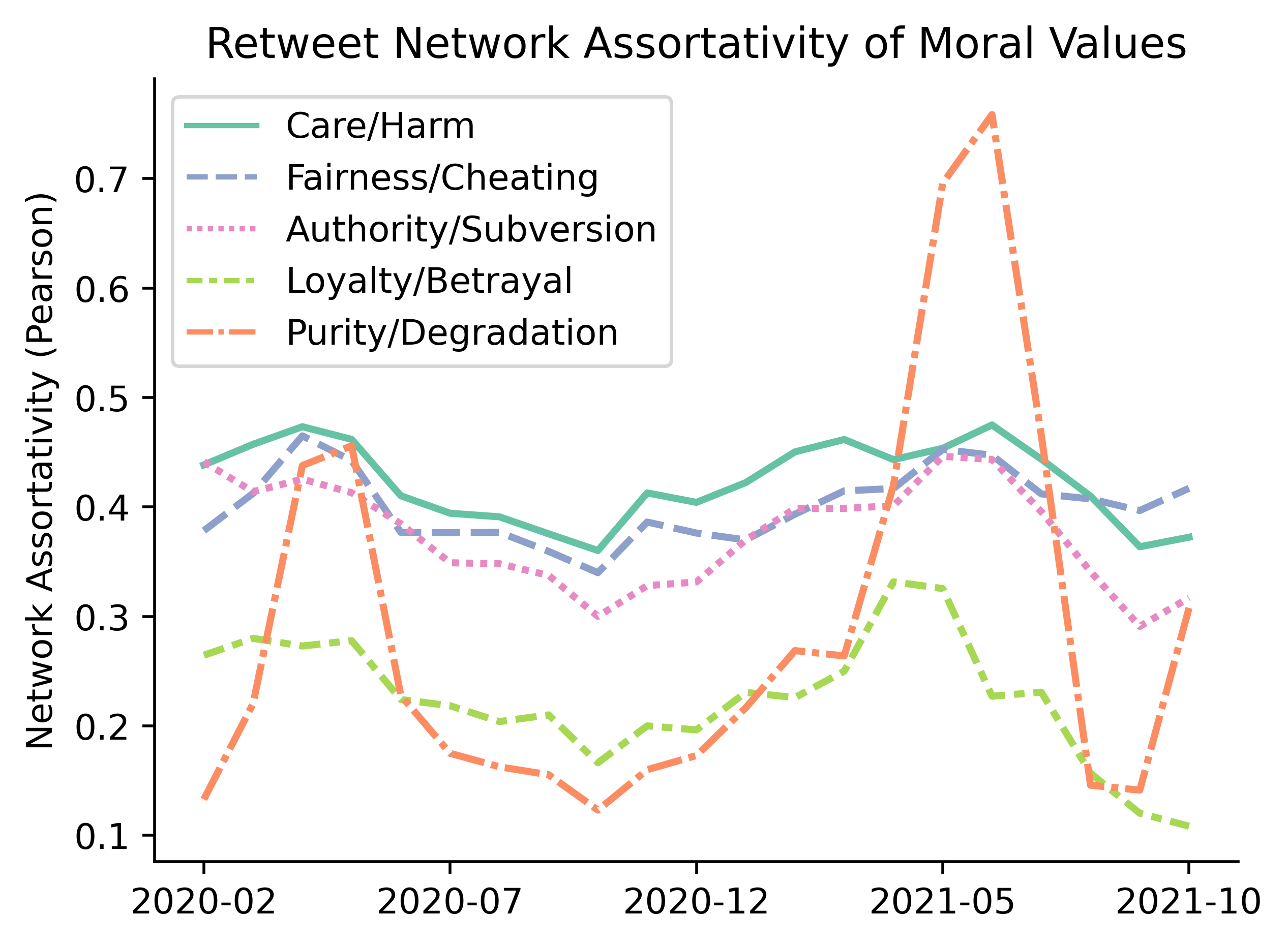}
    \caption{Retweet network assortativity of users' moral scores over time. High assortativity indicates homophily.}
    \label{fig:rt_assort_time}
\end{figure}

Figure \ref{fig:rt_assort_time} shows the results from the retweet networks, which are largely similar to the mention network. All of the Pearson correlation coefficients are positive and significant (all $p<0.001$), indicating moral assortative mixing or moral homophily. In particular, the \textit{care} $(\mu=0.43$), \textit{fairness} ($\mu=0.40$), and \textit{authority} ($\mu=0.38$) foundations indicate strong, consistent moral homophily. However, \textit{loyalty} $(\mu=0.22)$ reflects much lower assortativity. Intriguingly, while most foundations reflect consistent levels of homophily over time, the \textit{purity} foundation $(\mu=0.29)$ showcases fluctuations, at times recording both the lowest and highest homophily values. This variance can be attributed to increased discussions related to masking and vaccines, topics linked closely to the \textit{purity} foundation. Notably, peaks in \textit{purity} assortativity align with significant events, such as the CDC's official recommendation of face coverings in April 2020 and the widespread discussion of vaccines during the first half of 2021. While these findings align partially with prior research on \textit{purity} homophily \cite{dehghani2016purity}, it's crucial to highlight the consistent and strong homophily observed in other moral foundations, particularly \textit{care}, \textit{fairness}, and \textit{authority}.

\subsection{Homophily of User Groups}
Next, we evaluate whether there is homophily on the moral group level. We cannot use the assortativity method here because group membership is categorical, not numerical. However, we can empirically compute how often users from group $X$ will retweet or mention users from group $Y$ compared to a null model. Let $P(X\xleftarrow{rt} Y)$ be the actual proportion of tweets published by user group $Y$ that are retweeted by user group $X$ out of all the retweets by user group $X$. We then randomly re-assign the group identity of the users to compute the null model $P_{rand}(X\xleftarrow{rt} Y)$. This procedure controls for the fact that the moral groups are not even in size, so some user groups don't appear to receive more communication only because they have more users. We then examine $P/P_{rand}$, which would be greater than 1 if $X$ communicates with $Y$ more frequently than the null baseline and lower than 1 if $X$ communicates less frequently with $Y$. 

\begin{figure}
    \centering
    \includegraphics[width=0.8\linewidth]{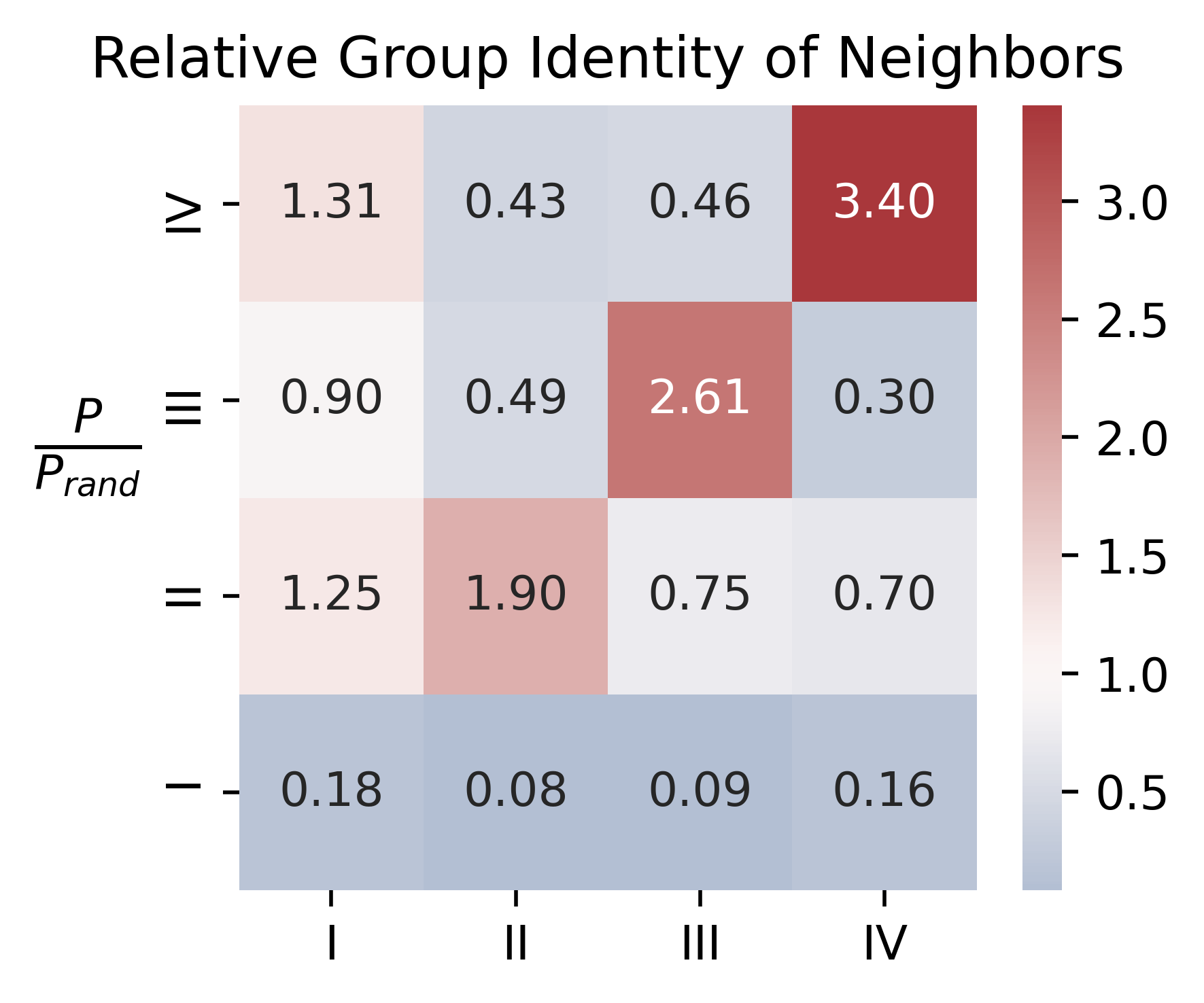}
    \caption{The ratio of how often users in group $X$ retweets users in group $Y$ divided by the null baseline amount. $>1$ (red) indicates $X$ is more likely to retweet from $Y$ than the null baseline. }
    \label{fig:rel_neighbor_group_rt}
\end{figure}
The results for the retweet interactions are shown in Figure \ref{fig:rel_neighbor_group_rt}. We omit the results for the mention interactions, which are very similar. Users in group IV (right-leaning users with \textit{fairness} and \textit{authority} morals) display the strongest homophily, preferentially retweeting from users in the same moral group more than 3x as much as the null baseline. This trend is followed by users in groups III (politically balanced users with \textit{care} and \textit{purity} morals) and II (politically balanced users with \textit{authority} and \textit{care} morals) but to a lesser extent. Users in groups II, III, and IV also retweet from the other three groups much less frequently than the null baseline. Interestingly, group I (left-leaning users with \textit{authority} and \textit{fairness} morals) users do not display preferential communication with in-group members. In fact, they retweeted themselves very infrequently and retweeted from other groups (II and IV) more than expected. As we saw in the visualization Figure \ref{fig:tsne}, this could indicate that group I users are not characterized by strong homophilous social network ties but rather serve as peripheral members of the Twittersphere interacting with all users.

\begin{figure}
    \centering
    \includegraphics[width=0.8\linewidth]{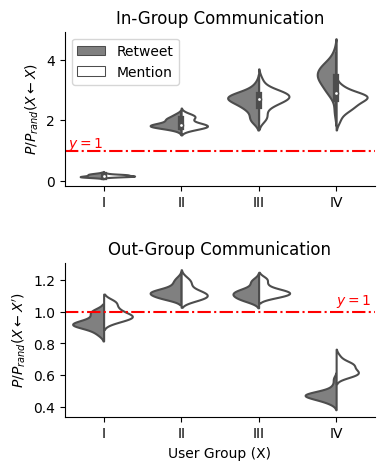}
    \caption{The ratio of how often users in group $X$ communicate with in-group users (top) or out-group users (bottom) divided by the null baseline amount. $>1$ indicates $X$ is more likely to retweet from $Y$ than the null baseline.}
    \label{fig:in_out_group}
\end{figure}

Figure \ref{fig:in_out_group} displays the same data but now comparing in-group communication $P/P_{rand}(X\leftarrow X)$ with out-group communication $P/P_{rand}(X\leftarrow X')$, where $X'$ include all the users not in $X$. We observe that group IV users have very high in-group communication compared to the null baseline, followed by groups III and II, whereas group I does not favor in-group communication. However, we also see that groups II and III, the politically balanced user groups, have high out-group communication compared to the null baseline. This is not true for group IV, which clearly prefers in-group communication and not out-group communication. This finding may signal that group IV, who are predominantly right-leaning extremists, is falling into a communication echo chamber, a potentially harmful manifestation of online communication. Our findings align with prior work considering only user political orientation, which showed that right-leaning users discussing COVID on Twitter are situated in a tight-knit political echo chamber \cite{jiang2021social}.

\section{RQ3: Bridging Moral Divides}

Previously, we found that users of certain moral typologies communicate frequently with in-group members (groups II, III, and IV), and some also display a lack of communication with out-group members (group IV). In this section, we consider whether there is any pattern in what morality of messages tends to work well traveling across group members. Here, we also want to contend with the fact that messages can contain more than one moral foundation. Let $m$ be the 5-dimensional multilabel indicator vector of the moral foundations present in a tweet. We then count the frequency of the moral foundation combinations of every tweet published by user group $X$ that was retweeted or mentioned by a user that is not from group $X$, denoted by $C(X, X', m)$. To draw a suitable comparison, we also count the moral foundation combinations' occurrences of retweets or mentions within a group, denoted by $C(X, X, m)$. Finally, we calculate the ratio $C(X, X',m)/C(X, X,m)$, which would reflect the moral foundation combinations that were retweeted more frequently by in-group members than out-group members. Sorting by this ratio, we identify the top five moral foundation combinations that were retweeted by out-group members for every group in Table \ref{tab:out_mf_combo}. We also highlight the moral foundations that are not the preferred moral foundation by that specific user group.

Key insights emerged from this analysis. Firstly, moral foundations that were less favored by the user group exhibited a notable trend of being retweeted more frequently by out-group members. Secondly, combinations featuring multiple moral foundations garnered heightened traction among out-group members, especially since tweets with multiple moral foundations are extremely rare: tweets with two moral foundations only make up 13\% of all tweets, and tweets with three moral foundations make up less than 1\% of all tweets. The implications are profound, suggesting avenues of facilitators of social network connections by enhancing one's moral diversity and moral plurality. We also offer support for research in moral reframing as a tool to bridge ideological gulfs \cite{feinberg2019moral}.

However, it's important to acknowledge the limitations of this analysis. While we shed light on existing communication patterns, we cannot definitively causal relationships. There might be untapped moral combinations that could potentially resonate well with out-group members but remain unexplored within our dataset. Nonetheless, these findings offer valuable insights into understanding the dynamics of moral communication using observable data.

\begin{table}
    \centering
    \begin{tabular}{llc}
         \toprule 
         \multicolumn{2}{l}{\textbf{Moral Foundations}} & Ratio \\
         \midrule
         \multicolumn{3}{l}{\textit{I: Authority \& Fairness}}\\
         & \textcolor{purple}{\textbf{Care}}, Purity & 2.316 \\
         & Fairness & 2.010 \\
         & \textcolor{purple}{\textbf{Care}}, Authority, Purity & 1.895 \\
         & \textcolor{purple}{\textbf{Care}}, Loyalty & 1.558 \\
         & \textcolor{purple}{\textbf{Care}}, Fairness, Loyalty & 1.293 \\
         \midrule 
         \multicolumn{3}{l}{\textit{II: Authority \& Care}}\\
        & Care, Loyalty, \textcolor{purple}{\textbf{Purity}} & 1.299 \\
        & \textcolor{purple}{\textbf{Fairness}}, Loyalty & 1.151 \\
        & Care, \textcolor{purple}{\textbf{Fairness}}, Loyalty & 1.137 \\
        & Authority, Loyalty, \textcolor{purple}{\textbf{Purity}} & 1.114 \\
        & Authority, Loyalty & 1.099 \\
         \midrule 
         \multicolumn{3}{l}{\textit{III: Care \& Purity}}\\
        & \textcolor{purple}{\textbf{Fairness}}, \textcolor{purple}{\textbf{Authority}}, Purity & 5.049 \\
        & Care, \textcolor{purple}{\textbf{Authority}}, Purity & 3.098 \\
        & Care, Purity & 2.059 \\
        & Care & 1.566 \\
        & \textcolor{purple}{\textbf{Fairness}}, \textcolor{purple}{\textbf{Authority}} & 1.177 \\
         
         \midrule 
         \multicolumn{3}{l}{\textit{IV: Fairness \& Authority}}\\
         & \textcolor{purple}{\textbf{Care}}, Loyalty & 4.053 \\
        & \textcolor{purple}{\textbf{Care}}, Authority, Loyalty & 2.303 \\
        & \textcolor{purple}{\textbf{Care}}, Authority, \textcolor{purple}{\textbf{Purity}}& 1.822 \\
        & \textcolor{purple}{\textbf{Care}}, Loyalty & 1.614 \\
        & Authority & 1.158 \\
        \bottomrule 
    \end{tabular}
    \caption{For every user group, we show the top five moral combinations used in messages that are retweeted more often by out-group members than in-group members. The list is sorted by the ratio $ C(X, X',m)/C(X, X,m)$. Highlighted moral foundations are the ones that the user group does not favor (cf. Figure \ref{fig:user_clusters}). }
    \label{tab:out_mf_combo}
\end{table}

\section{Discussion}

In this paper, we present a large-scale empirical investigation of COVID-19 online conversation through the lens of moral psychology. Using a large dataset of COVID-19 tweets spanning nearly two years, we offer insights into the characteristics of the main types of users from a moral standpoint (RQ1), communication patterns among users of different moral typologies (RQ2), and how the morality of messages leads to more effective diverse communication.

Based on users' moral foundation profiles, user metadata features, and social network data, we uncover four distinct groups of users that differ in moral preferences, strength, and political affiliations. Group I users represent low-influence, left-leaning users who care mainly about \textit{authority} and \textit{fairness}, positioning themselves mostly at the peripherals of social network communication. Group II users is a politically balanced group who care about \textit{authority} and \textit{Care}. The biggest contrast occurs between group III, who highly value \textit{care} and \textit{purity}, and group IV, who highly value \textit{fairness} and \textit{authority}. Additionally, groups III and IV value the exact moral foundations that the other one does not. The two groups also differ considerably in their political partisanship composition; group III users are politically balanced while group IV users are predominantly far-right. 

Analyzing communication patterns based on morality, our results illustrate patterns of moral homophily. We consistently find high homophily in the \textit{care}, \textit{fairness}, and \textit{authority} foundations. We also partially align our results with established findings on \textit{purity} homophily \cite{dehghani2016purity}, although we observe \textit{purity} homophily only at specific time points that seem to be contextually related. Regarding group identity homophily, Groups II, III, and IV display increased communication within their respective groups. Notably, Group IV not only demonstrates heightened in-group communication but also a significant lack of communication with out-group members, hinting at the potential existence of harmful moral echo chambers. These findings provide valuable insights into the dynamics of moral homophily, emphasizing its presence across different moral foundations and its potential impact on group communication dynamics.

Finally, upon analyzing the moral foundation of messages that could bridge moral divides, we find that moral diversity and moral pluralism may be useful approaches. Messages that contain moral foundations their users don't usually prefer, and messages that contain multiple moral foundations tend to be more likely retweeted by out-group members.

The findings of our study offer crucial insights and recommendations for practitioners, health agencies, and researchers. We emphasize significant variations among users concerning moral preferences, suggesting that both the occurrence of communication and the content of messages are influenced by users' moral typology. Notably, we identify group IV users, characterized by a preference for the \textit{fairness} and \textit{authority} foundations and primarily conservative in nature, as potentially more susceptible to morally homogeneous messages. Adding to the wealth of literature on moral ideals separating the political left and right \cite{haidt2007morality,graham2009liberals}, we observe that the moral lines are not always so clear cut. Users on both sides of the political spectrum can prioritize the same moralities that are both individualizing (\textit{care} or \textit{fairness}) and binding (\textit{authority} or \textit{purity}). A deeper dive into the complexity of our moral differences, in addition to political ideology factors, can lead to a more comprehensive understanding of online communication. Finally, our research sheds light on the potential usefulness of rhetorical tools focused on moral reframing to enhance the diversity of communication on online platforms.

\subsection{Limitations}
The research conducted in this study is strictly observational, and no causal relationships can be implied from the findings. The validity of the results is contingent upon the accuracy of various machine learning models utilized in the study, including those for morality detection and political partisanship detection. Additionally, it's essential to note that the scope of the results is limited to the topic of COVID, and generalizability to other topics may not be assured. These methodological considerations and limitations should be taken into account when interpreting and applying the study's results.

\subsection{Ethical Statement}
The data used in our study is publicly accessible \cite{chen2020covid}. Our study was exempt from review by the Institutional Review Board (IRB) as it solely relied on publicly available data. During our analysis, we protect user privacy by utilizing user IDs instead of screen names. Further, in the interest of user confidentiality, we present only aggregated statistics in this paper. In conclusion, we put forth ethical recommendations, proposing the integration of moral psychology in comparable research endeavors and the development of rhetorical tools specifically designed for moral reframing. This initiative aims to enrich the diversity of communication on online platforms. Acknowledging a potential risk, we recognize the possibility of malicious actors manipulating public opinion by manipulating moral foundations. However, we contend that the positive outcomes and contributions of our research far outweigh this risk. The authors declare no competing interests.

\subsection{Acknowledgment}
This work was supported by DARPA (award number HR001121C0169).

\bibliography{aaai22}

\begin{thebibliography}{44}
\providecommand{\natexlab}[1]{#1}

\bibitem[{Beir{\'o} et~al.(2023)Beir{\'o}, D'Ignazi, Perez~Bustos, Prado, and Kalimeri}]{beiro2023moral}
Beir{\'o}, M.~G.; D'Ignazi, J.; Perez~Bustos, V.; Prado, M.~F.; and Kalimeri, K. 2023.
\newblock Moral Narratives Around the Vaccination Debate on Facebook.
\newblock In \emph{The ACM Web Conference 2023}, 4134--4141.

\bibitem[{Borghouts et~al.(2023)Borghouts, Huang, Gibbs, Hopfer, Li, and Mark}]{borghouts2023understanding}
Borghouts, J.; Huang, Y.; Gibbs, S.; Hopfer, S.; Li, C.; and Mark, G. 2023.
\newblock Understanding Underlying Moral Values and Language Use of COVID-19 Vaccine Attitudes on Twitter.
\newblock \emph{PNAS Nexus}, 2(3): pgad013.

\bibitem[{Boyd, Golder, and Lotan(2010)}]{boyd2010tweet}
Boyd, D.; Golder, S.; and Lotan, G. 2010.
\newblock Tweet, Tweet, Retweet: Conversational Aspects of Retweeting on Twitter.
\newblock In \emph{HICSS 2010}, 1--10. IEEE.

\bibitem[{Bruchmann and LaPierre(2022)}]{bruchmann2022moral}
Bruchmann, K.; and LaPierre, L. 2022.
\newblock Moral Foundations Predict Perceptions of Moral Permissibility of COVID-19 Public Health Guideline Violations in United States University Students.
\newblock \emph{Frontiers in Psychology}, 12: 6442.

\bibitem[{Chan(2021)}]{chan2021moral}
Chan, E.~Y. 2021.
\newblock Moral Foundations Underlying Behavioral Compliance During the COVID-19 Pandemic.
\newblock \emph{Personality and Individual Differences}, 171: 110463.

\bibitem[{Chen, Lerman, and Ferrara(2020{\natexlab{a}})}]{chen2020tracking}
Chen, E.; Lerman, K.; and Ferrara, E. 2020{\natexlab{a}}.
\newblock Tracking Social Media Discourse About the COVID-19 Pandemic: Development of a Public Coronavirus Twitter Data Set.
\newblock \emph{JMIR Public Health and Surveillance}, 6(2): e19273.

\bibitem[{Chen, Lerman, and Ferrara(2020{\natexlab{b}})}]{chen2020covid}
Chen, E.; Lerman, K.; and Ferrara, E. 2020{\natexlab{b}}.
\newblock Tracking Social Media Discourse About the {COVID-19} Pandemic: Development of a Public {Coronavirus} {Twitter} Data Set.
\newblock \emph{JMIR Public Health Surveill}, 6(2): e19273.

\bibitem[{Dehghani et~al.(2016)Dehghani, Johnson, Hoover, Sagi, Garten, Parmar, Vaisey, Iliev, and Graham}]{dehghani2016purity}
Dehghani, M.; Johnson, K.; Hoover, J.; Sagi, E.; Garten, J.; Parmar, N.~J.; Vaisey, S.; Iliev, R.; and Graham, J. 2016.
\newblock Purity Homophily in Social Networks.
\newblock \emph{Journal of Experimental Psychology: General}, 145(3): 366.

\bibitem[{Devlin et~al.(2019)Devlin, Chang, Lee, and Toutanova}]{bert}
Devlin, J.; Chang, M.-W.; Lee, K.; and Toutanova, K. 2019.
\newblock {BERT}: Pre-training of Deep Bidirectional Transformers for Language Understanding.
\newblock In \emph{NAACL-HLT 2019}, 4171--4186.

\bibitem[{D{\'\i}az and Cova(2022)}]{diaz2022reactance}
D{\'\i}az, R.; and Cova, F. 2022.
\newblock Reactance, Morality, and Disgust: The Relationship Between Affective Dispositions and Compliance With Official Health Recommendations During the COVID-19 Pandemic.
\newblock \emph{Cognition and Emotion}, 36(1): 120--136.

\bibitem[{Ekici, Y{\"u}cel, and Cesur(2021)}]{ekici2021deciding}
Ekici, H.; Y{\"u}cel, E.; and Cesur, S. 2021.
\newblock Deciding Between Moral Priorities and COVID-19 Avoiding Behaviors: A Moral Foundations Vignette Study.
\newblock \emph{Current Psychology}, 1--17.

\bibitem[{Feinberg and Willer(2019)}]{feinberg2019moral}
Feinberg, M.; and Willer, R. 2019.
\newblock Moral Reframing: A Technique for Effective and Persuasive Communication Across Political Divides.
\newblock \emph{Social and Personality Psychology Compass}, 13(12): e12501.

\bibitem[{Graham et~al.(2013)Graham, Haidt, Koleva, Motyl, Iyer, Wojcik, and Ditto}]{graham2013moral}
Graham, J.; Haidt, J.; Koleva, S.; Motyl, M.; Iyer, R.; Wojcik, S.~P.; and Ditto, P.~H. 2013.
\newblock Moral Foundations Theory: The Pragmatic Validity of Moral Pluralism.
\newblock In \emph{Advances in Experimental Social Psychology}, volume~47, 55--130. Elsevier.

\bibitem[{Graham, Haidt, and Nosek(2009)}]{graham2009liberals}
Graham, J.; Haidt, J.; and Nosek, B.~A. 2009.
\newblock Liberals and Conservatives Rely on Different Sets of Moral Foundations.
\newblock \emph{Journal of Personality and Social Psychology}, 96(5): 1029.

\bibitem[{Graham, Nosek, and Haidt(2012)}]{graham2012moral}
Graham, J.; Nosek, B.~A.; and Haidt, J. 2012.
\newblock The Moral Stereotypes of Liberals and Conservatives: Exaggeration of Differences Across the Political Spectrum.
\newblock \emph{PloS ONE}, 7(12): e50092.

\bibitem[{Gramlich(2022)}]{pew2022normal}
Gramlich, J. 2022.
\newblock Two Years Into the Pandemic, Americans Inch Closer to a New Normal.
\newblock \emph{Pew Research Center}.
\newblock \url{https://www.pewresearch.org/2022/03/03/two-years-into-the-pandemic-americans-inch-closer-to-a-new-normal/} Accessed: April 4, 2023.

\bibitem[{Guo, Mokhberian, and Lerman(2023)}]{guo2023data}
Guo, S.; Mokhberian, N.; and Lerman, K. 2023.
\newblock A Data Fusion Framework for Multi-Domain Morality Learning.
\newblock In \emph{ICWSM 2017}, 281--291.

\bibitem[{Haidt(2012)}]{haidt2012righteous}
Haidt, J. 2012.
\newblock \emph{The Righteous Mind: Why Good People Are Divided by Politics and Religion}.
\newblock Vintage.

\bibitem[{Haidt and Graham(2007)}]{haidt2007morality}
Haidt, J.; and Graham, J. 2007.
\newblock When Morality Opposes Justice: Conservatives Have Moral Intuitions that Liberals May Not Recognize.
\newblock \emph{Social Justice Research}, 20(1): 98--116.

\bibitem[{Haidt and Joseph(2004)}]{haidt2004intuitive}
Haidt, J.; and Joseph, C. 2004.
\newblock Intuitive Ethics: How Innately Prepared Intuitions Generate Culturally Variable Virtues.
\newblock \emph{Daedalus}, 133(4): 55--66.

\bibitem[{Hatemi, Crabtree, and Smith(2019)}]{hatemi2019ideology}
Hatemi, P.~K.; Crabtree, C.; and Smith, K.~B. 2019.
\newblock Ideology Justifies Morality: Political Beliefs Predict Moral Foundations.
\newblock \emph{American Journal of Political Science}, 63(4): 788--806.

\bibitem[{Hemsley et~al.(2018)Hemsley, Stromer-Galley, Semaan, and Tanupabrungsun}]{hemsley2018tweeting}
Hemsley, J.; Stromer-Galley, J.; Semaan, B.; and Tanupabrungsun, S. 2018.
\newblock Tweeting to the Target: Candidates’ Use of Strategic Messages and @Mentions on Twitter.
\newblock \emph{Journal of Information Technology \& Politics}, 15(1): 3--18.

\bibitem[{Jiang et~al.(2020)Jiang, Chen, Yan, Lerman, and Ferrara}]{jiang2020political}
Jiang, J.; Chen, E.; Yan, S.; Lerman, K.; and Ferrara, E. 2020.
\newblock Political Polarization Drives Online Conversations About {COVID-19} in the United States.
\newblock \emph{Human Behavior Emerging Technologies}, 2(3): 200--211.

\bibitem[{Jiang and Ferrara(2023)}]{jiang2023socialllm}
Jiang, J.; and Ferrara, E. 2023.
\newblock Social-LLM: Modeling User Behavior at Scale using Language Models and Social Network Data.
\newblock \emph{arXiv preprint arXiv:2401.00893}.

\bibitem[{Jiang, Ren, and Ferrara(2021)}]{jiang2021social}
Jiang, J.; Ren, X.; and Ferrara, E. 2021.
\newblock Social Media Polarization and Echo Chambers in the Context of COVID-19: Case Study.
\newblock \emph{JMIRx Med}, 2(3): e29570.

\bibitem[{Jiang, Ren, and Ferrara(2023)}]{retweetbert}
Jiang, J.; Ren, X.; and Ferrara, E. 2023.
\newblock Retweet-BERT: Political Leaning Detection Using Language Features and Information Diffusion on Social Networks.
\newblock In \emph{ICWSM 2023}, 459--469.

\bibitem[{Johnson and Goldwasser(2018)}]{johnson2018classification}
Johnson, K.; and Goldwasser, D. 2018.
\newblock Classification of Moral Foundations in Microblog Political Discourse.
\newblock In \emph{ACL 2018}, 720--730.

\bibitem[{Kaplan et~al.(2023)Kaplan, Vaccaro, Henning, and Christov-Moore}]{kaplan2023moral}
Kaplan, J.~T.; Vaccaro, A.; Henning, M.; and Christov-Moore, L. 2023.
\newblock Moral Reframing of Messages About Mask-Wearing During the COVID-19 Pandemic.
\newblock \emph{Scientific Reports}, 13(1): 10140.

\bibitem[{Koleva et~al.(2012)Koleva, Graham, Iyer, Ditto, and Haidt}]{koleva2012tracing}
Koleva, S.~P.; Graham, J.; Iyer, R.; Ditto, P.~H.; and Haidt, J. 2012.
\newblock Tracing the Threads: How Five Moral Concerns (Especially Purity) Help Explain Culture War Attitudes.
\newblock \emph{Journal of Research in Personality}, 46(2): 184--194.

\bibitem[{Kossinets and Watts(2009)}]{kossinets2009origins}
Kossinets, G.; and Watts, D.~J. 2009.
\newblock Origins of Homophily in an Evolving Social Network.
\newblock \emph{American Journal of Sociology}, 115(2): 405--450.

\bibitem[{{Los Angeles Times}(2020)}]{latimes2020}
{Los Angeles Times}. 2020.
\newblock Letters to the Editor: Refusing To Wear a Mask Is the Most Brainless, Selfish Way To Assert Your Liberty.
\newblock \emph{Los Angeles Times}.
\newblock \url{https://www.latimes.com/opinion/story/2020-06-20/refusing-to-wear-a-mask-brainless-selfish} Accessed: April 4, 2023.

\bibitem[{Luttrell and Trentadue(2023)}]{luttrell2023advocating}
Luttrell, A.; and Trentadue, J.~T. 2023.
\newblock Advocating for Mask-Wearing Across the Aisle: Applying Moral Reframing in Health Communication.
\newblock \emph{Health Communication}, 1--13.

\bibitem[{McPherson, Smith-Lovin, and Cook(2001)}]{mcpherson2001birds}
McPherson, M.; Smith-Lovin, L.; and Cook, J.~M. 2001.
\newblock Birds of a Feather: Homophily in Social Networks.
\newblock \emph{Annual Review of Sociology}, 27(1): 415--444.

\bibitem[{Mejova, Kalimeri, and Morales(2023)}]{mejova2023authority}
Mejova, Y.; Kalimeri, K.; and Morales, G. D.~F. 2023.
\newblock Authority Without Care: Moral Values Behind the Mask Mandate Response.
\newblock In \emph{ICWSM 2023}, 614--625.

\bibitem[{Metaxas et~al.(2015)Metaxas, Mustafaraj, Wong, Zeng, O'Keefe, and Finn}]{metaxas2015retweets}
Metaxas, P.; Mustafaraj, E.; Wong, K.; Zeng, L.; O'Keefe, M.; and Finn, S. 2015.
\newblock What do Retweets Indicate? Results From User Survey and Meta-Review of Research.
\newblock In \emph{ICWSM 2015}, volume~9, 658--661.

\bibitem[{Newman(2003)}]{assortativity}
Newman, M.~E. 2003.
\newblock Mixing Patterns in Networks.
\newblock \emph{Physical Review E}, 67(2): 026126.

\bibitem[{O'Hara and Stevens(2015)}]{o2015echo}
O'Hara, K.; and Stevens, D. 2015.
\newblock Echo Chambers and Online Radicalism: Assessing the Internet's Complicity in Violent Extremism.
\newblock \emph{Policy \& Internet}, 7(4): 401--422.

\bibitem[{Pacheco et~al.(2022)Pacheco, Islam, Mahajan, Shor, Yin, Ungar, and Goldwasser}]{pachec2022holistic}
Pacheco, M.~L.; Islam, T.; Mahajan, M.; Shor, A.; Yin, M.; Ungar, L.; and Goldwasser, D. 2022.
\newblock A Holistic Framework for Analyzing the {COVID}-19 Vaccine Debate.
\newblock In \emph{NAACL-HLT 2022}, 5821--5839.

\bibitem[{Rao et~al.(2023)Rao, Guo, Wang, Morstatter, and Lerman}]{rao2023pandemic}
Rao, A.; Guo, S.; Wang, S.-Y.~N.; Morstatter, F.; and Lerman, K. 2023.
\newblock Pandemic Culture Wars: Partisan Asymmetries in the Moral Language of COVID-19 Discussions.
\newblock \emph{arXiv preprint arXiv:2305.18533}.

\bibitem[{Reimer et~al.(2022)Reimer, Atari, Karimi-Malekabadi, Trager, Kennedy, Graham, and Dehghani}]{reimer2022moral}
Reimer, N.~K.; Atari, M.; Karimi-Malekabadi, F.; Trager, J.; Kennedy, B.; Graham, J.; and Dehghani, M. 2022.
\newblock Moral Values Predict County-Level COVID-19 Vaccination Rates in the United States.
\newblock \emph{American Psychologist}, 77(6): 743.

\bibitem[{Rojecki, Zheleva, and Levine(2021)}]{rojecki2021moral}
Rojecki, A.; Zheleva, E.; and Levine, L. 2021.
\newblock The Moral Imperatives of Self-Quarantining.
\newblock In \emph{Annual Meeting of the American Political Science Association}.

\bibitem[{Schmidtke et~al.(2022)Schmidtke, Kudrna, Noufaily, Stallard, Skrybant, Russell, and Clarke}]{schmidtke2022evaluating}
Schmidtke, K.~A.; Kudrna, L.; Noufaily, A.; Stallard, N.; Skrybant, M.; Russell, S.; and Clarke, A. 2022.
\newblock Evaluating the Relationship Between Moral Values and Vaccine Hesitancy in Great Britain During the COVID-19 Pandemic: A Cross-Sectional Survey.
\newblock \emph{Social Science \& Medicine}, 308: 115218.

\bibitem[{Singh et~al.(2021)Singh, Kaur, Matsuo, Iyengar, and Sasahara}]{singh2021morality}
Singh, M.; Kaur, R.; Matsuo, A.; Iyengar, S.; and Sasahara, K. 2021.
\newblock Morality-Based Assertion and Homophily on Social Media: A Cultural Comparison Between English and Japanese Languages.
\newblock \emph{Frontiers in Psychology}, 12: 768856.

\bibitem[{Van~der Maaten and Hinton(2008)}]{tsne}
Van~der Maaten, L.; and Hinton, G. 2008.
\newblock Visualizing Data Using t-SNE.
\newblock \emph{Journal of Machine Learning Research}, 9(11).

\end{thebibliography}

\end{document}